\documentclass[osajnl,twocolumn,showpacs]{revtex4}
\usepackage{graphicx}
\graphicspath{{pict/}{}}

\usepackage{bm}%

\newcounter{Fig}

\newcommand{\be}{\begin{equation}}
\newcommand{\ee}{\end{equation}}

\begin{document}

\title{Surface solitons in two-dimensional quadratic photonic lattices}

\author{Mario I. Molina$^1$ and Yuri S. Kivshar$^2$}

\affiliation{$^1$Departmento de F\'{\i}sica, Facultad de Ciencias,
Universidad de Chile, Santiago, Chile\\
$^2$Nonlinear Physics Center, Research School of Physics
and Engineering, Australian National University, Canberra ACT 0200,
Australia}


\begin{abstract}
We study two-color surface solitons in two-dimensional photonic lattices with quadratic nonlinear response.
We demonstrate that such parametrically coupled optical localized modes can exist in the corners or at the edges
of a square photonic lattice, and we analyze the impact of the phase mismatch on their properties, stability, 
and the threshold power for their generation.
\end{abstract}


\maketitle

Two-dimensional surface solitons have been recently predicted to exist
as novel types of discrete solitons localized in the corners or at the edges of
two-dimensional photonic lattices~\cite{makris_2D,pla_our,pre_2D}.
These theoretical predictions were followed by the experimental observation of
two-dimensional surface solitons in optically-induced photonic
lattices~\cite{prl_1} and waveguide arrays
laser-written in fused silica~\cite{prl_2}. Importantly, these
two-dimensional surface solitons demonstrate novel features
in comparison with their counterparts in truncated one-dimensional
waveguide arrays. In particular, in a sharp contrast to one-dimensional
surface solitons, the threshold power of two-dimensional surface solitons
is lower at the surface than in a bulk making the mode excitation easier~\cite{pla_our}.

Surface solitons are usually considered for cubic or saturable nonlinear media.
However, multicolour discrete solitons in quadratically nonlinear lattices have been studied
theoretically in both one- and two-dimensional lattices~\cite{chi2_p1,chi2_p2,chi2_p3,chi2_p4}
irrespective to the surface localization effects. Only Siviloglou et al.~\cite{OE_stegeman} studied
discrete quadratic surface solitons experimentally in periodically poled lithium
niobate waveguide arrays, and they employed a discrete model with decoupled waveguides
at the second harmonics to model some of the effects observed experimentally.

More elaborated theory of one-dimensional surface solitons in truncated quadratically nonlinear photonic
lattices, the so-called two-color surface lattice solitons, has been developed recently by
Xu and Kivshar~\cite{chi2_our} who analyzed the impact of the phase mismatch on the existence
and stability of nonlinear parametrically coupled surface modes, and also found novel classes of 
one-dimensional two-color twisted surface solitons which are stable in a large domain of their existence.

In this Letter, we extend the analysis of two-color surface solitons to the case
of two-dimensional photonic lattices. We study, for the first time to our knowledge, two-color surface solitons in two-dimensional square photonic lattices with quadratic nonlinear response.
We analyze the effect of mismatch on the existence, stability, and generation
of surface solitons located in the corners or at the edges of the nonlinear lattice.

We consider the propagation of light in a two-dimensional
photonic lattice of a finite extent imprinted in a quadratic nonlinear medium, which involves
the interaction between the fundamental frequency (FF) and
second-harmonic (SH) waves. Light propagation is
described by the following coupled nonlinear discrete equations~\cite{chi2_p1,chi2_our}
\begin{eqnarray}
\label{eq:model}
&& i\frac{d u_{n,m}}{d
z}+C_u \Delta_2 u_{n,m} +2\gamma
u_{n,m}^{*}v_{n,m} \exp(+i\beta z)=0, \nonumber \\
&& i\frac{d v_{n,m}}{d
z} + C_v \Delta_2 v_{n,m}
+\gamma u_{n,m}^2 \exp(-i\beta z)=0,
\end{eqnarray}
where $u_{n,m}$ and $v_{n,m}$ are the normalized amplitudes of the FF and SH waves, respectively,
$C_u$ and $C_v$ are the coupling coefficients,
$\gamma$ characterizes the second-order nonlinearity, and $\beta$ is the effective mismatch between two harmonics.
The second-order difference operator $\Delta_2$ is defined as
\[\Delta_2u_{n,m} = \sum_{i=\pm1}(u_{n+i,m} + u_{n,m+i}).\]
Equations (\ref{eq:model}) conserve the total power, 
\begin{equation}
P=P_u +P_v=\sum_{n,m}(|u_{n,m}|^2 + 2|v_{n,m}|^2).
\label{power}
\end{equation}

We look for stationary two-mode solutions of Eq.~(\ref{eq:model}) in the form, $u_{n,m}(z) = U_{n,m} \exp(i \lambda z)$ and
$v_{n,m}(z) = V_{n,m} \exp(2i \lambda z - i \beta z)$, and obtain the nonlinear algebraic equations for the (real)
mode amplitudes, $U_{n,m}$ and $V_{n,m}$,
\begin{eqnarray}
-\lambda U_{n,m} + C_{u}\Delta_2 U_{n,m} +2 \gamma U_{n,m} V_{n,m} & =& 0,\nonumber\\
 -2 \lambda V_{n,m} + C_{v}\Delta_2 V_{n,m}+\beta V_{n,m}+\gamma U_{n,m}^{2} &= &0,
 \label{eq:2}
\end{eqnarray}
where for the $(N$x$M)$ lattice we have $n=0,1,\ldots, N$, $m=0,1,\ldots,M$ and $U_{n,m}=0$ and $V_{n,m}=0$ if either $n<0$, or $m<0$, or $n>N$, or $m>M$.

Equations (\ref{eq:2}) possess symmetry properties. For instance, the transformation $\gamma\rightarrow -\gamma, V_{n,m}\rightarrow -V_{n,m}$ leave Eq.~(\ref{eq:2}) invariant.

\begin{figure}[h]
\centering
\includegraphics[width=4.2cm]{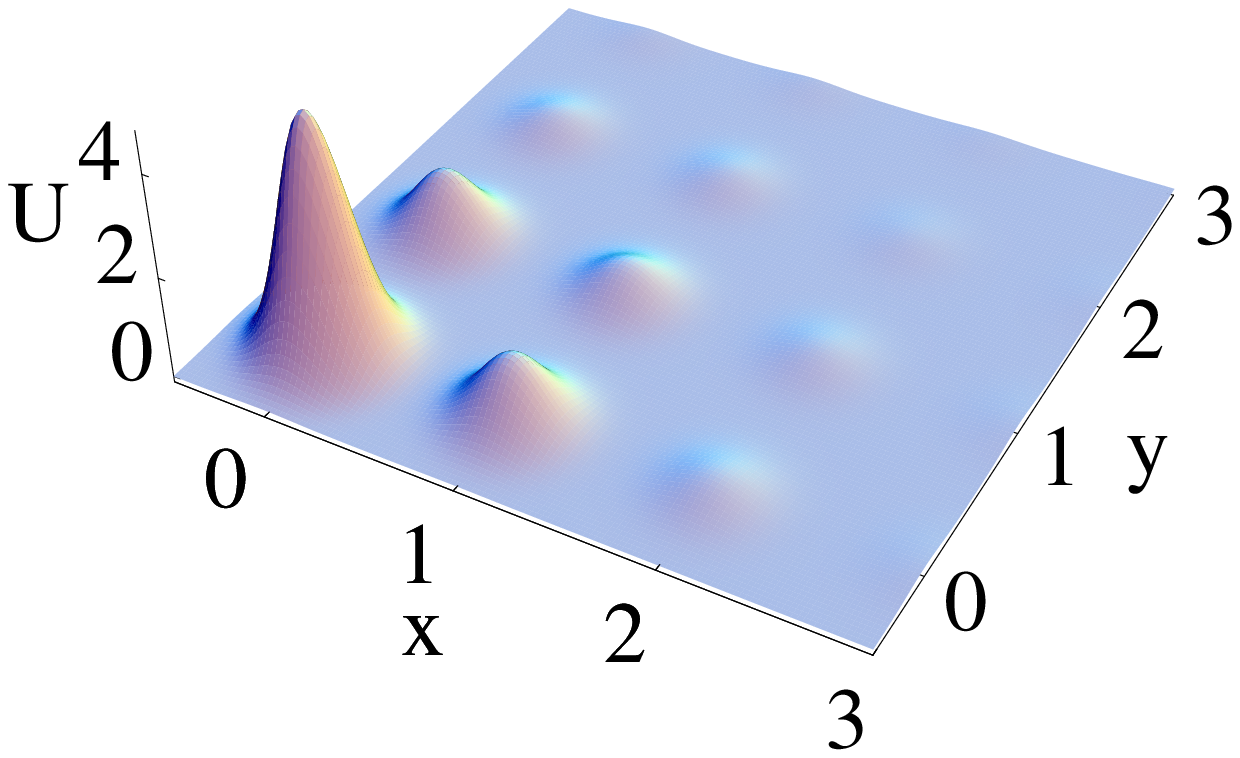}
\includegraphics[width=4.2cm]{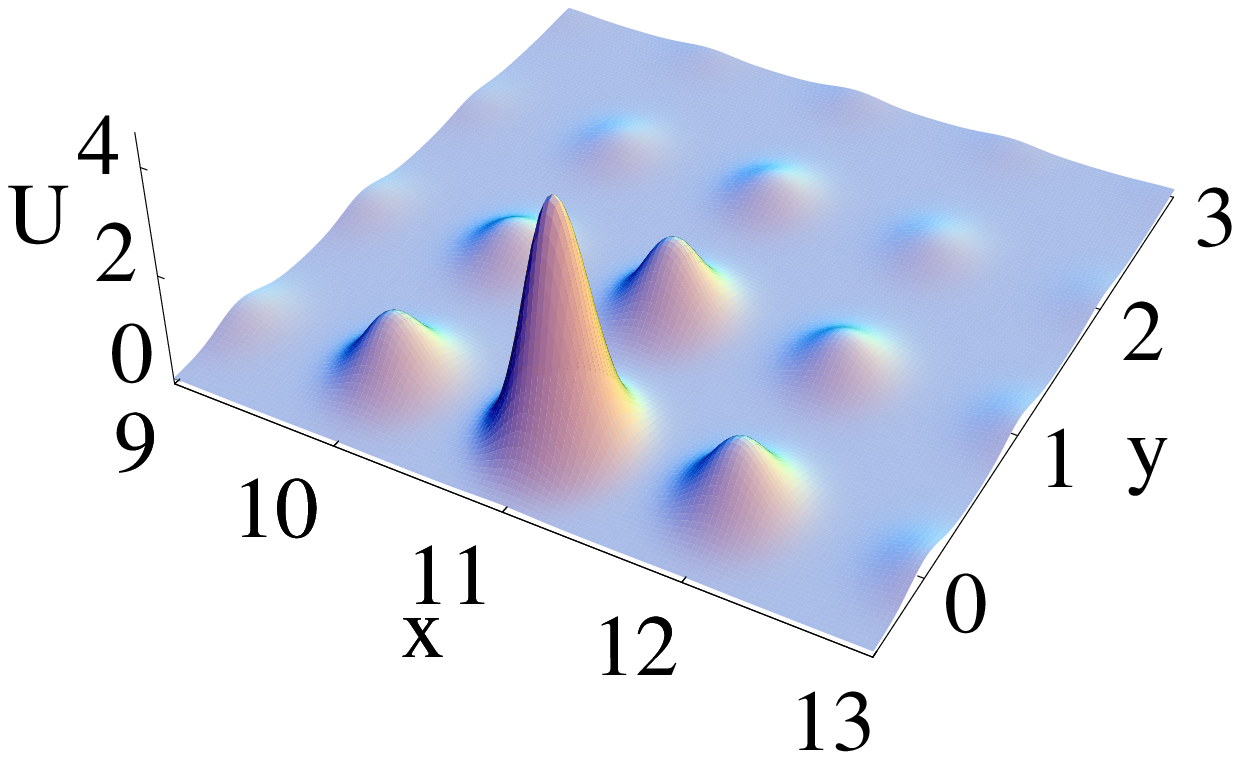}
\caption{(color online) Examples of the corner (left) and edge (right) localized modes in two-dimensional quadratic
lattice. Shown are the amplitudes of the FF components 
for $C_u = 1$, $C_v = 0.5$, $\beta = 0$,
and $\lambda =5$.}
\label{fig1}
\end{figure}
In the anti-continuum limit, i.e. when the couplings in the lattice vanish,
Eqs.~(\ref{eq:2}) imply $V_{n,m} = \lambda/2 \gamma$ and
$U_{n,m}^{2}=(\lambda/2\gamma)(2 \lambda-\beta)/\gamma$. Thus, if $\lambda>0$, then
$2 \lambda>\beta$. If $\lambda<0$, then $2 \lambda<\beta$. Also, localized modes should exist outside the linear spectrum band, $|\lambda|> 4 C_{u}$. Thus, in the propagation constant-mismatch space,  the region where localized modes exist is bounded by $\lambda>4 C_{u}$, $\lambda>\beta/2$, or $\lambda<-4 C_{u}$, $\lambda<\beta/2$. Also in this limit, the propagation constant is proportional to $P^{2}$ rather than $P$, as in the case of the cubic nonlinearity.

We consider a square lattice of a finite extent with $N=M=21$, and look for two-dimensional localized modes in
the corner ($1,1$) and at the edge ($N/2,1$) of the lattice.  We use two sets of the coupling parameters: (i) strong coupling, $C_{u}=1$ and $C_{v}=0.5$, this is similar to the parametric processes in bulk media, and (b) no coupling for the second harmonic, $C_{u}=1$ and $C_{v}=0$, similar to the fabricated structures employed in the recent experiments~\cite{OE_stegeman}. For given values of the coupling parameters $(C_{u}, C_{v})$, nonlinear parameter $\gamma$, and mismatch $\beta$, we use a standard numerical procedure of continuation from the anti-continuum limit. Examples of the nonlinear corner and edge nonlinear localized states are shown in Fig.~\ref{fig1}, where the fields propagating along the waveguides are presented as a superposition of the waveguide modes, $U(x,y)=\sum_{n,m} U_{n,m} \phi(x-n,y-m)$, where $\phi(x,y)$ is the (single mode) guide centered on site $(n,m)$, and similarly for $V(x,y)$. In Figs. 1 and 4, we use $\phi(x,y)=\exp(-x^2-y^2)/\sigma^2$, with $\sigma^2=0.05$.

\begin{figure}[h]
\centering
\includegraphics[width=4.2cm]{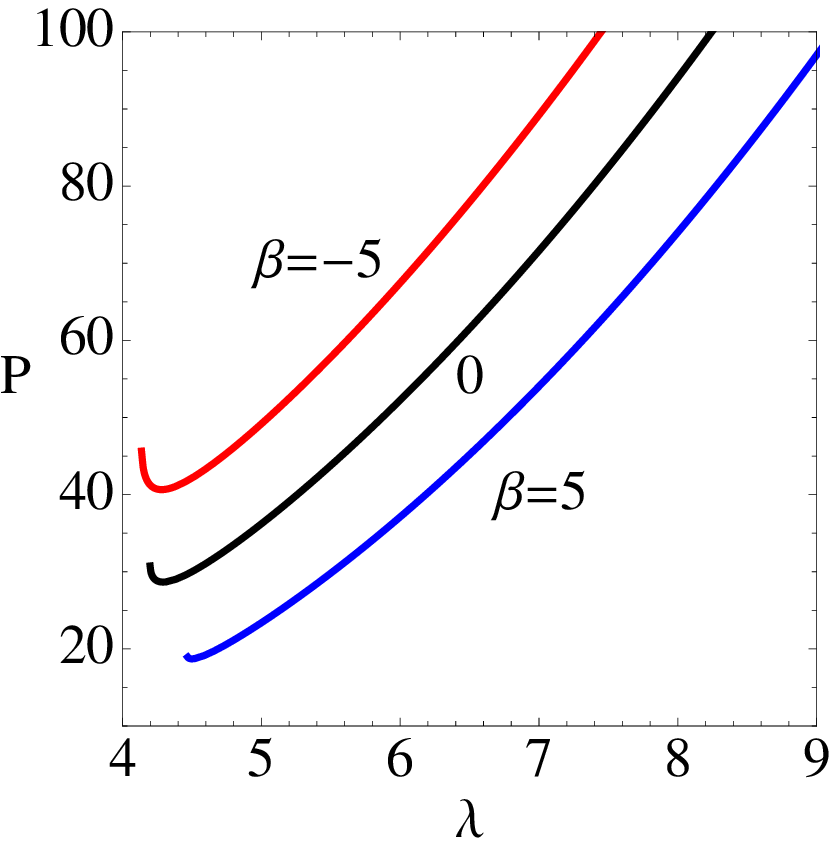}
\includegraphics[width=4.2cm]{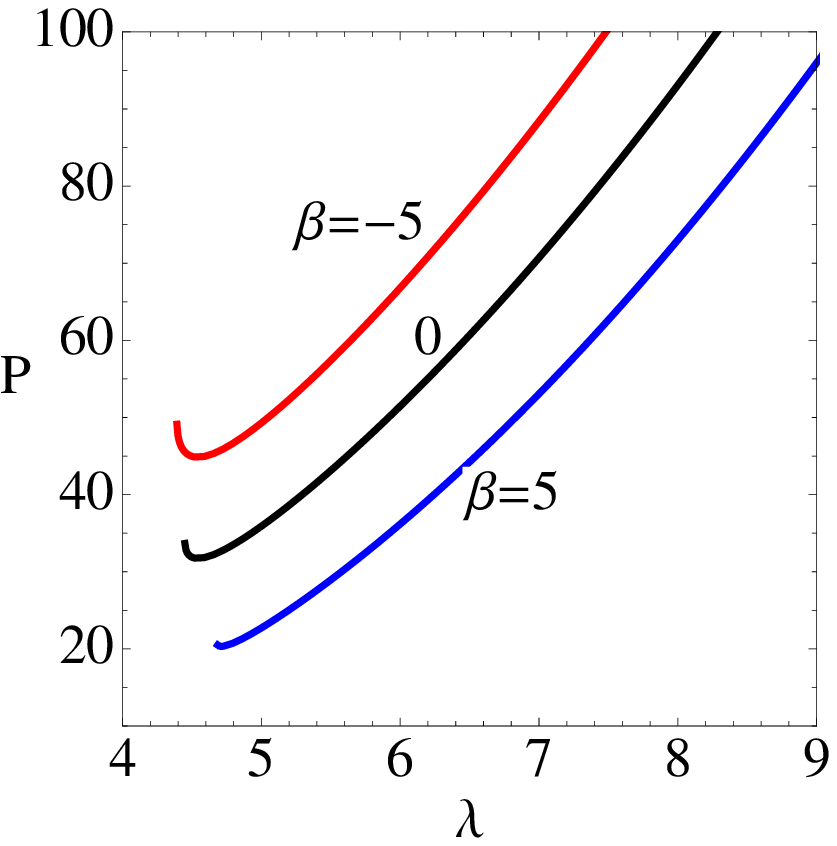}
\caption{(color online) Total power of the two-dimensional modes localized in the corner (left) and at the
edge (right) of the lattice for $C_u = 1$, $C_v = 0.5$,
and for three different values of the mismatch parameter $\beta$.}
\label{fig2}
\end{figure}

We focus on the analysis of the power dependencies characterizing the families of two-dimensional localized modes. The typical dependencies of the total power (\ref{power}) on the propagation constant $\lambda$ are shown in Fig.~\ref{fig2}, for several values of the mismatch parameter $\beta$ marked on the plots. The power curve has a typical minimum corresponding to the minimum (threshold) power for the localized mode to exist. In the region where the slope of the power $P$ becomes negative, the localized modes are unstable, as observed for other types of surface modes and also verified here by numerical analysis.

\begin{figure}[h]
\centering
\includegraphics[width=4.2cm]{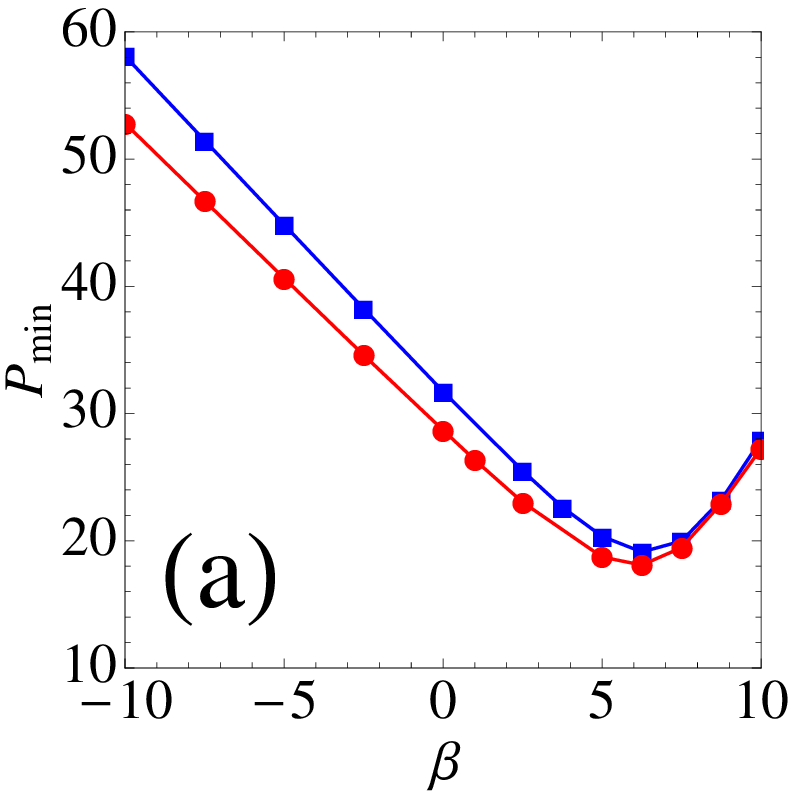}
\includegraphics[width=4.2cm]{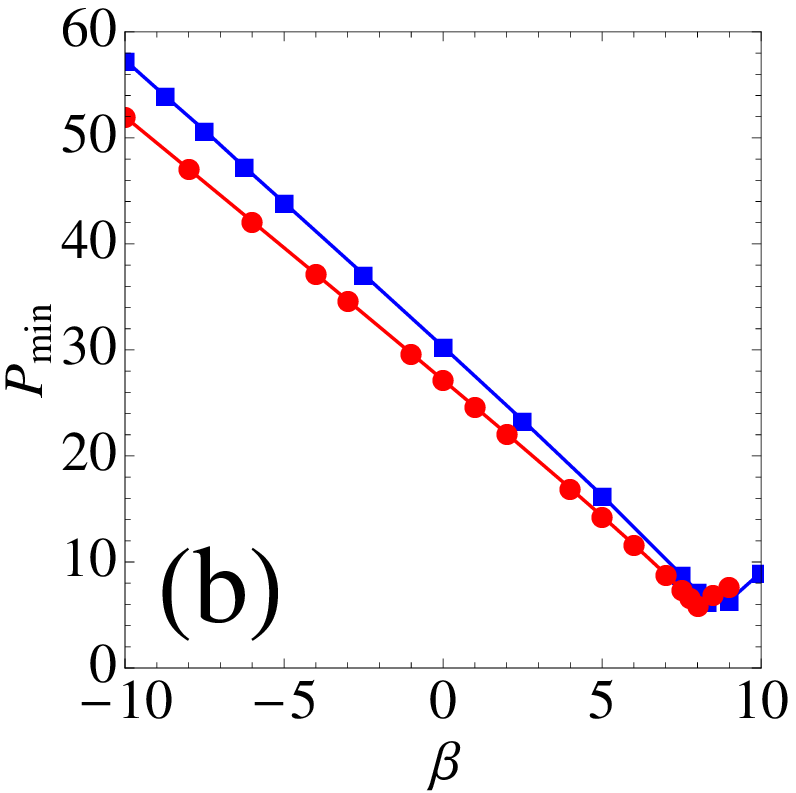}
\caption{(color online) Minimum power vs. mismatch for the corner (circle) and edge (square) modes, for
(a) $C_u=1$, $C_v=0.5$ and (b) $C_u=1$, $C_v=0$. For virtually all values of the mismatch
parameter, the corner mode requires less power than the edge mode.}
\label{fig3}
\end{figure}

Figure~\ref{fig3} shows the dependence of the minimum power $P_{min}$ required to create a stable localized mode vs. mismatch $\beta$, either in the corner or at the edge of the lattice. In general, the results for the corner and edge modes
are rather similar. Clearly, the value of the mismatch  determines the threshold power needed to generate a localized state, with a minimum value observed for positive $\beta$ larger than $\beta=5$. The minimum power depends also upon the value of the coupling constants. However, the main result is that for virtually all values of the mismatch parameter $\beta$, the corner mode requires less power than the edge mode; this is somewhat similar to the case of the cubic nonlinear lattice~\cite{makris_2D,pla_our,pre_2D}.

\begin{figure}[h]
\centering
\includegraphics[width=4.2cm]{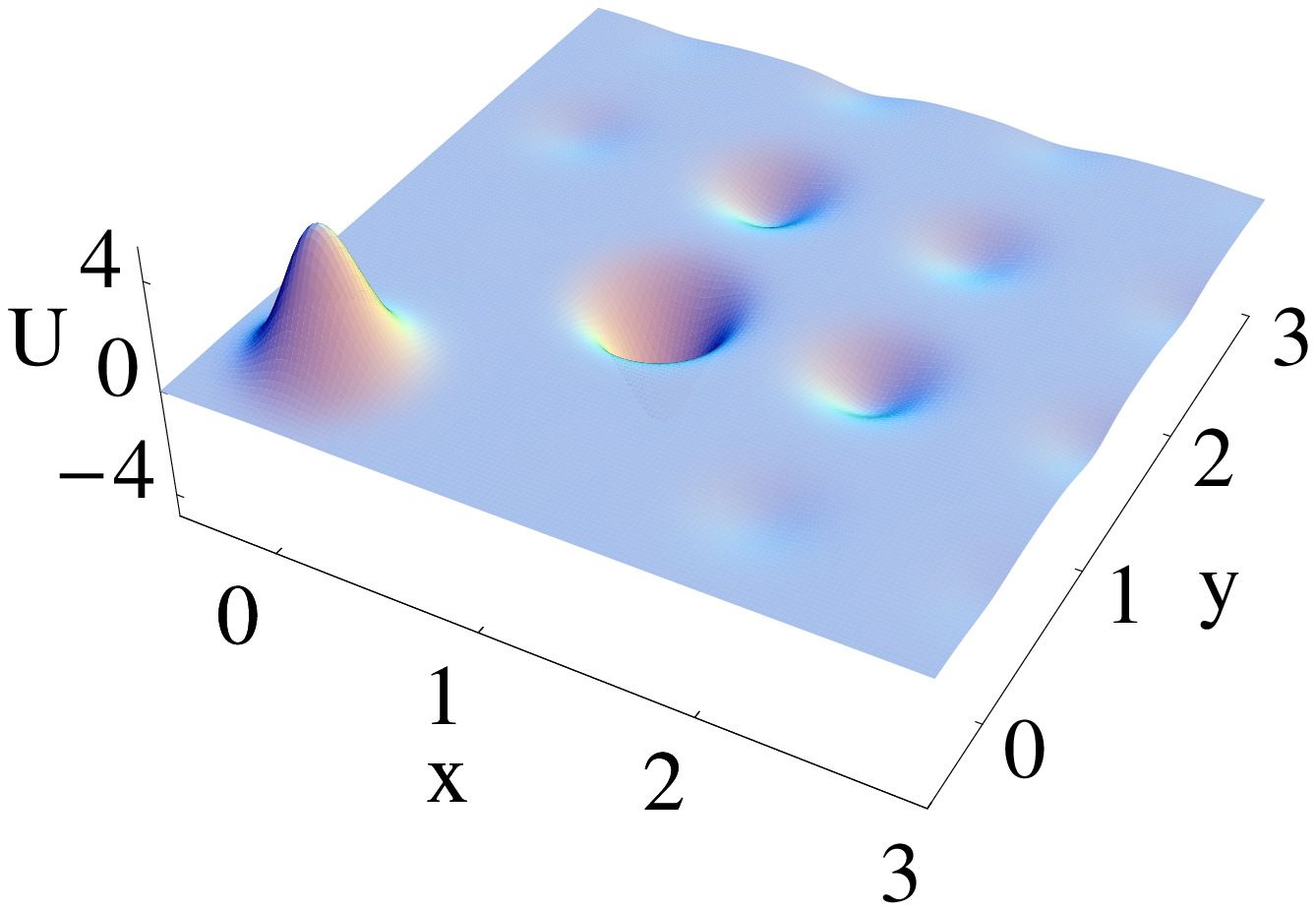}
\includegraphics[width=4.2cm]{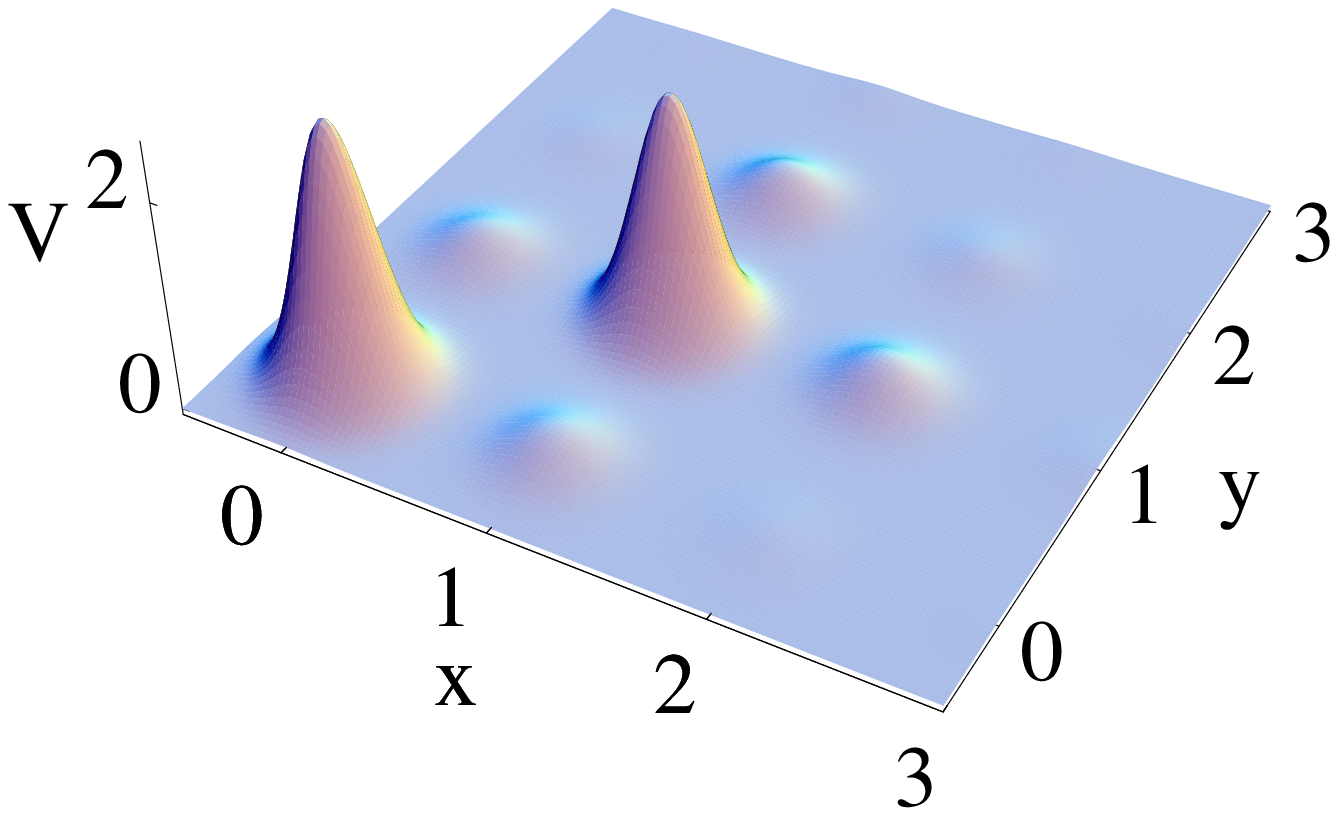}
\caption{(color online) Example of a two-dimensional twisted mode at the lattice corner.
Shown are the amplitude of the FF (left) and SH (right) fields for $C_u = 1$, $C_v = 0.5$, $\lambda=5$, and $\beta=0$.}
\label{fig4}
\end{figure}

Using the same numerical approach and starting from the anti-continuum limit, we find several other families of two-dimensional localized modes located in a close vicinity of the lattice corners and edges. These modes
provide a two-dimensional generalization of the surface modes known for one-dimensional lattices placed at different distances from the edge, and corresponding to a crossover between the surface and bulk discrete solitons as discussed earlier~\cite{OL_molina}. In addition, we find novel classes of the so-called two-dimensional twisted modes, an example of one of such modes is shown in Fig.~\ref{fig4} for $C_u = 1$, $C_v = 0.5$, and $\lambda=5$.

To study the generation of these two-color surface modes, we launch a tight beam at one site,
$u_{n,m}(0) = U_0\delta_{n,n_0}\delta_{m,m_0}$ and $v_{n,m}(0)= V_0\delta_{n,n_0}\delta_{m,m_0}$,
either in the corner or at the edge of the lattice. We let the system evolve from $z = 0$ up to $z = z_{\rm max}$, and trace the shape of the beam measuring the partial powers $P_u$ and $P_v$ for both FF and SH components, respectively. In general, the results are qualitatively similar for the two initial beam positions (in the corner and at the edge). The only difference is in the actual value of the field amplitude at which the beam self-trapping occurs. On the other hand, significant differences are observed between two sets of the coupling parameters and also
whether both fields or only the FF field is initially present.

\begin{figure}[h]
\centering
\includegraphics[width=8.8cm]{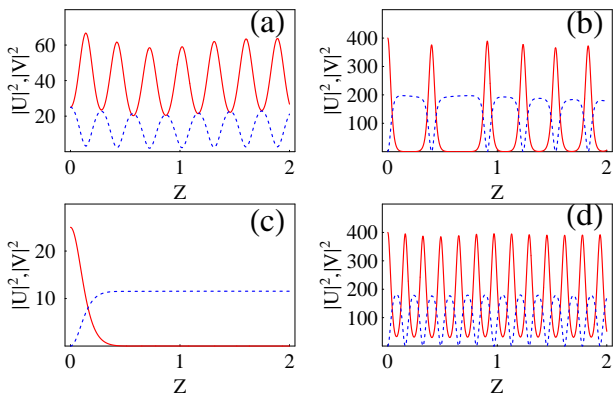}
\caption{(color online) Excitation of two-color surface modes at the lattice corner. Shown is the dynamics of FF (solid) and SH (dashed) components for (a,b) $C_u = 1$, $C_v = 0.5$; $\beta = 0$
with (a) $U_0= V_0=5$ and (b) $U_0 = 20$, $V_0 = 0$; and (c,d) $C_u = 1$, $C_v = 0$, with (c) $U_0 = 5$, $V_0 = 0$, $\beta¯ = 0$, and $U_0 = 20$, $V_0 = 0$, $\beta= -5$.}
\label{fig5}
\end{figure}

In general, when both the fields are initially excited and the power is strong enough, a localized surface mode
is formed, whose amplitude oscillates at a relatively fast rate [see Fig.~\ref{fig5}(a)]. Presence of
a mismatch tends to destroy the localization, but it is restored if the initial power is high enough.
When the SH field is initially absent, we observe that it is harder
to excite a localized mode [see Fig.~\ref{fig5}(b)]. When it is achieved, the mode always
seem to oscillate between zero and a maximum amplitude at a fast rate.
When the SH coupling is absent ($C_v = 0$) and the harmonics are matched ($\beta= 0$), it can be proven that the
resulting equation for the FF field is transformed into a  discrete equation with a nonlinear loss term:\ $i (d/dz)u_{n,m}+C_{u} \Delta_{2} u_{n,m}+ 2 i \gamma^{2} u_{n,m}^{*} \int_{0}^{z} u_{n,m}^{2}(s) ds=0$. This induces a quick power loss in the FF field and its rapid transfer to the SH field [see Fig.~\ref{fig5}(c)]. The addition of mismatch and a high enough initial FF field, restores self-trapping for both the fields, where both the fields oscillate at a fast rate [see Fig.~\ref{fig5}(d)].

In conclusion, we have studied localization of light in two-dimensional quadratically nonlinear photonic lattices and determined the conditions for the existence of two-color surface states localized in the corners or at the edges of the lattice. We have analyzed the impact of the phase mismatch on the properties and stability of two-dimensional localized modes, as well as the threshold power for their generation.

This work was supported by Fondecyt grant 1080374 and by the
Australian Research Council.


\begin{thebibliography}{10}

\bibitem{makris_2D} K.G. Makris, J. Hudock, D.N. Christodoulides, G.
Stegeman, O. Manela, and M. Segev, Opt. Lett. {\bf 31}, 2774 (2006).

\bibitem{pla_our} R.A. Vicencio, S. Flach, M.I. Molina, and Yu.S.
Kivshar, Phys. Lett. A {\bf 364}, 274 (2007).

\bibitem{pre_2D} H. Susanto, P.G. Kevrekidis, B.A. Malomed, R.
Carretero-Gonz\'alez, and D.J. Franzeskakis, Phys. Rev. E {\bf 75}, 056605 (2007).

\bibitem{prl_1} X. Wang, A. Bezryadina, Z. Chen, K.G. Makris, D.N.
Christodoulides, and G.I. Stegeman, Phys. Rev. Lett. {\bf 98},
123903 (2007).

\bibitem{prl_2} A. Szameit, Y.V. Kartashov, F. Dreisow, T. Pertsch,
S. Nolte, A. T\"unnermann, and L. Torner,
Phys. Rev. Lett. {\bf 98}, 173903 (2007).

\bibitem{chi2_p1}  B.A. Malomed, P.G. Kevrekidis, D.J. Franzeskakis, H.E. Nistazakis,
and A.N. Yannacopoulos, Phys. Rev. E {\bf 65}, 056606 (2002).

\bibitem{chi2_p2} Ya.V. Kartashov, L. Torner, and V.A. Vysloukh, Opt. Lett. {\bf 29}, 1117 (2004).

\bibitem{chi2_p3} Z. Xu, Ya.V. Kartashov, L.-C. Crasovan, D. Mihalache, and L. Torner,
Phys. Rev. E {\bf 71}, 016616 (2005).

\bibitem{chi2_p4} H. Susano, P.G. Kevrekidis, R. Carretero-Gonzalez, B.A. Malomed,
and D.J. Franzeskakis, Phys. Rev. Lett. {\bf 99}, 214103 (2007).

\bibitem{OE_stegeman} G.A. Siviloglou, K.G. Makris, R. Iwanow, R. Schiek, D.N. Christodoulides,
G.I. Stegeman, Y. Ming, and W. Sohler, Opt. Express {\bf 14}, 5508 (2006).

\bibitem{chi2_our} Z. Xu and Yu.S. Kivshar, Opt. Lett. {\bf 33}, 2551 (2008).

\bibitem{OL_molina}
M. Molina, R. Vicencio, and Yu.~S. Kivshar, Opt. Lett. {\bf 31},
1693 (2006).

\end{thebibliography}
\end{document}